\documentclass[12pt]{article}
\usepackage{hyperref}
\usepackage{amsmath,amsthm,amssymb,bigdelim,multirow}
\usepackage{latexsym, enumerate}
\usepackage{amsfonts}
\usepackage[dvips]{graphicx} 

\numberwithin{equation}{section}

\theoremstyle{theorem}
\newtheorem{thm}{Theorem}
\newtheorem{prop}[thm]{Proposition}

\newtheorem{rem}[thm]{Remark}
\date{}
\title{\bf Calibration of Transparency Risks: \\
a Note\thanks{This research is partially supported 
by Open Research Center Project 
for Private Universities: matching fund subsidy from MEXT, 2004-2008 
and also by Grants-in-Aids for Scientific Research (No. 18540146) 
from the Japan Society for Promotion of Sciences.}}

\author{Jir\^o Akahori\thanks{Department of Mathematical Sciences and Research Center for Finance, 
Ritsumeikan University, 
1-1-1 Nojihigashi, Kusatsu, Shiga, 525-8577, Japan. 
E-mail: {\tt akahori@se.ritsumei.ac.jp}}, 
Yuuki Kanishi\thanks{Resona Trust \& Banking Co. Ltd
E-mail: {\tt yuuki.77.kanishi@hotmail.co.jp}} 
and Yuichi Morimura\thanks{Mitsubishi UFJ Securities, 
E-mail: {\tt yuing001@gmail.com}
}}

\begin{document}
\maketitle
\begin{abstract}
The aim of this research is to give
a simple framework to evaluate/quantize 
the {\em transparency} of a firm.
We assume that 
the process of the firm value is 
only observable once in a while  
but is strongly correlated with
the stock price which is observable and tradable. 
This hybrid type structure make the transparency 
``observable".  The implication of the present 
study is that the depth of the shock to the market 
caused by the precise accounting information 
does reflect 
the degree of transparency. 
Furthermore, it can be quantized resorting to 
the calibration method. 
\end{abstract}

\section{Introduction}

\subsection{Credit Risk Modeling: Literature}
The models of credit risks 
are often classified into two groups
by the degree of details in 
modeling how the default occurs.
In the {\em reduced form} models
such as Jarrow-Turnbull's \cite{JT2} 
and Duffie-Singleton's \cite{DuSi},   
the default is an exogenous event,
and its probabilistic nature is directly modeled 
in terms of the default probability
or the hazard rate, etc. 
On the other hand, 
if the default is endogenous/set to be  
a consequence of 
some economic activities, 
the model is called {\em structural}. 
In most structural models including
Merton \cite{M} and Leland-Toft \cite{LT}, 
the default occurs
when the firm value process reaches a boundary. 

As Jarrow and Protter \cite{JP} pointed out, 
the two approaches can be unified 
by introducing asymmetry between the 
the manager's perspective and the market's.   
The pioneering work in this hybrid approach was done by 
Duffie and Lando \cite{DL}. 
They assumed a default structure of Leland-Toft type 
which is not directly observable to the market. 
This filtering approach of Duffie-Land was generalized into the 
continuous time framework 
by Kusuoka \cite{Ku} and Nakagawa \cite{Na}. 

\subsection{Hybrid Models and Transparency Risk}
It is noteworthy that in the hybrid approach 
the degree of accounting transparency
is modeled implicitly by 
the difference of the manager's perspective 
and the market's one.
In fact, 
it is claimed in \cite[p.634 l.30--32]{DL} that 
{\em the shape of the term structure
of credit spreads may indeed play a useful empirical role in estimating the
degree of transparency of a firm}, and 
in \cite[p.134 l.13--15]{Na} it is pointed out that
{\em the credit spread can be explained by the hazard rate plus
some fluctuation caused by the difference} between the manager's
filtration and the market's one. That is to say, we have the 
following decomposition:
\begin{equation}\label{decomp}
\text{credit risk} = \text{default risk} + \text{transparency risk}.
\end{equation}
With this view, we can say that in the structural approach 
only default risk is modeled while 
in the reduced form approach credit risk 
is treated without decomposing. The transparency risk is made visible 
only when one construct a hybrid type model. 

\subsection{Our Motivation/Calibration}
The importance of the accounting 
transparency is now widely recognized 
among the managers as well as the investors. 
Of course the Sarbanes-Oxley (SOx) Act of 2002 in U.S.A, 
followed by J-SOx of September 2007 in Japan, 
was a cornerstone but 
they are basically aimed to reduce 
the transparency risks from the investor's perspective. 
The decomposition (\ref{decomp}), however, 
suggests that the managers can reduce the credit spread 
by promoting transparency. It says there can be 
positive incentives for the manager to reduce the transparency risks. 
This observation is empirically supported by the study of Yu
\cite{Yu}. 

The problem is then how we could know the degree of 
transparency. The formula of Duffie-Lando or 
the one by Nakagawa is too complicated to calibrate 
it to the market values. We need simpler formulas but 
nonetheless it should be based on a hybrid type model. 

\subsection{Structure of Transparency/
Slightly Incomplete Market}
Motivated by the above demand, in the present paper 
we will construct a simple hybrid type model 
out of the classical Merton's structural model. 
Structure of transparency is modeled by 
``$ \rho $-coupling" of Wiener processes.
Roughly speaking, the filtration of manager's 
and market's are generated by two $ 1 $-dimensional
Wiener processes $ W $ and $ W' $ starting 
from $ 0 $ respectively.
Here 
\begin{equation}\label{rho}
 \langle W, W' \rangle_t = \rho t 
\end{equation}
for a constant 
$ \rho \in (0,1) $, which is set  
the unique parameter describing the transparency. 

We will work on the continuous time framework but 
the full accounting information (or the firm value) is supposed to be 
available at the discrete set of 
dates $ t_1,\ldots ,t_n,\ldots $. 
These discrete filtration make the market ``slightly incomplete"
and thus we need to be careful about the consistency 
with the no-arbitrage framework. 

\subsection{Main Results}
Under the simple hybrid model assumptions described roughly in the 
above, the present study will show that
\begin{itemize}
\item with a proper choice of the state price density, 
the market value of a firm is obtained 
by the image of the Ornstein-Uhlenbeck semigroup,  
\item the credit spread formula is explicitly obtained 
and it doesn't depend explicitly on the transparency parameter 
at the dates $ t_1, \ldots, t_n, \ldots $, 
\item and by this property the parameter can be 
calibrated to the market value.  
\end{itemize}  

\subsection{Organization of the Rest of the Paper}
In section \ref{Setting}, we shall give the setting and 
the market model on which we will be working. 
The first result on the consistency of 
our market model is given as Theorem \ref{firstresult}. 
In section \ref{Econ}, presented is a result on an economic property 
of our market model. 
Under these settings, we will obtain a generic formula (\ref{CDt}) 
in section \ref{Generic}. The formula is applied to evaluation of 
the credit risk and transparency risk 
in sections \ref{DS} and \ref{Calib}.
Proofs for those mathematical results are given in 
sections \ref{pf-first}, \ref{pf-vo}, and \ref{pf-eqg}.

\section{The Formulas}
\subsection{The Firm Value and the Market}\label{Setting}
Let $(\Omega,\mathcal{F},\mathbf{P})$ be a probability space
on which a one dimensional Wiener process
$ \{ W_t \}$ can be defined, and 
$ \{ \mathcal{F}_t \} $ be the natural filtration of $W$.
We assume Merton's economy in \cite{M}; i.e, 
the firm value 
is assumed to be 
\begin{eqnarray}\label{FV1}
V_t := V_0 \exp((\mu - \frac{\sigma^2}{2})t + \sigma W_t),
\end{eqnarray}
where $\mu \in \mathbf{R}$ and $\sigma \geq 0$. 

Let $ \mathbf{T} := \{ t_k : k \in \mathbf{Z}_+ \} \subset \mathbf{R}_+ $,
where $ t_0 = 0 $ and $ t_k < t_{k+1} $ for each $ k \in \mathbf{Z}_+ $, 
be the dates of the accounting report; namely 
we suppose that the firm value $ V $ 
is observable only at each $t_k \in \mathbf{T} $,
and during the interval $(t_k, t_{k+1})$
the market can only ``guess" the firm value. 
Let 
\begin{eqnarray*}
\mathcal{G}_t := \mathcal{F}'_t \vee
\sigma (V_s : s \in \mathbf{T}, s \leq t ), 
\end{eqnarray*}
where 
$\mathcal{F}'_t$ is the natural filtration of $W'$ 
which is  
introduced in Introduction. Note that 
the joint law of $ W $ and $ W' $
is completely determined by (\ref{rho}). 

To be consistent with no-arbitrage framework, 
we set the state price density of the market
by 
\begin{equation*}
Z_t = 
\exp ( \theta W'_t - \frac{1}{2} \theta^2 t  - r t ), 
\end{equation*}
where $ r $ is the constant interest rate and 
\begin{equation*}
\theta := -\frac{\mu - r}{\sigma \rho}. 
\end{equation*}

It should be noted that 
$ Z_t e^{rt} $ is a martingale with respect to 
the filtration $ \{ \mathcal{G} \} $
and therefore a probability measure $ \mathbf{Q} $ on
$ \{ \mathcal{G} \} $ is defined by
\begin{equation*}
\frac{d \mathbf{Q}}{d \mathbf{P}} \bigg|_{\mathcal{G}_t }
= Z_t e^{rt}.
\end{equation*}

Our first result is the following. 
\begin{thm}\label{firstresult} 
The ``filtered" firm value process 
$ V'_t := \mathbf{E}[V_t| \mathcal{G}_t] $
defines the no-arbitrage price
with respect to the state price density $ Z $ defined 
above. Namely, 
for $t_{k}\leq  t< t_{k+1}$, we have
\begin{equation*}
V'_t =
Z_t^{-1} \mathbf{E} [ Z_{t_{k+1} } V_{t_{k+1}} | \mathcal{G}_t ]
= e^{-r (t_{k+1} - t)} \mathbf{E}^\mathbf{Q} 
[V_{t_{k+1}} | \mathcal{G}_t ]. 
\end{equation*}
In particular, $ V'_t e^{-rt} $ is a $ \mathbf{Q} $-$ \mathcal{G} $-
martingale. 
\end{thm}
A proof of Theorem \ref{firstresult}
will be given in section \ref{pf-first}. 

\begin{rem}{\em 
The market value of the firm $ V' $ is explicitly
given as 
\begin{equation}\label{V'}
\begin{split}
V'_t &= V_{t_{k}}\exp((\mu - \frac{(\sigma \rho)^2}{2})(t-t_k)\\
& + \sigma \rho (W'_t - W'_{t_k})),
\end{split}
\end{equation}
which is the image of Ornstein-Uhlenbeck Semigroup $ T_u $
defined by setting $ u = \log \rho $. 
At the dates $t_k$, 
the market value of the firm $V'_{t_k}$ 
coincides with the firm value $V_{t_k}$.
When $\rho=1$, the market fully observe 
the firm value even during the period $(t_k,t_{k+1})$. 
}
\end{rem}

\subsection{An Economic Interpretation 
of the State Price Density $ Z $}\label{Econ}

Since we set $ Z $ to be a state price density, 
\begin{equation*}
C(X)_t = Z_t^{-1} \mathbf{E} [ Z_{t_{k+1} } X | \mathcal{G}_t ]
= e^{-r (t_k - t)} \mathbf{E}^\mathbf{Q} 
[X | \mathcal{G}_t ]. 
\end{equation*}
gives a fair value at time $ t  (< t_k )$
of the cash flow $ X $ at time 
$ t_k \in \mathbf{T} $.
The following proposition could give 
an economic interpretation of the process
$ C(X) $. 
\begin{prop}\label{vo}
The value $ C(X) $ is the replication cost (within the market
$ \mathcal{G} $ ) of
\begin{eqnarray*}
K^* := \mathbf{E}[ X | \mathcal{G}_{t_k -}],
\end{eqnarray*}
which is minimizer of
\begin{eqnarray*}
\inf\{||K_T - X||_{L^2} : K_T 
\in L^2(\mathcal{G}_{t_k -})\}, 
\end{eqnarray*}
where 
\begin{eqnarray*}
\mathcal{G}_{t-} :=  \vee_{s < t} \mathcal{G}_s.
\end{eqnarray*}
Namely we have, for $t<t_k$
\begin{equation*}
C(X)_t =  e^{-r(t_k-t)} \mathbf{E}^\mathbf{Q}[K^* | \mathcal{G}_t]
=  e^{-r(t_k-t)} \mathbf{E}^\mathbf{Q}[X | \mathcal{G}_t] .
\end{equation*}
\end{prop}
A proof will be given in section \ref{pf-vo}.

\subsection{The Generic Formulas}\label{Generic}
In this section 
we will obtain explicit formulas 
for $ C(X) $ with $ X =f(V_{t_n})$
for a bounded Borel function 
\begin{eqnarray*}
f:\mathbf{R} \rightarrow \mathbf{R}. 
\end{eqnarray*}

Now, let
\begin{equation*}
\begin{split}
\tau^{\sigma, \rho} (s) 
&:=-\frac{\sigma^2}{2} (t_n - t_{k}) \\
&+ \frac{(\sigma \rho)^2}{2}(s - t_{k})+ r(t_n-s),
\end{split}
\end{equation*}
\begin{equation*}
\nu^{\sigma, \rho} (s)
:=\sigma \sqrt{ (t_n-t_{k})- \rho^2(s -t_{k})}.
\end{equation*}

\begin{thm}\label{eqg}
We have the following explicit formulas for
the value $C(f(V_{t_n}) )$:
\begin{equation}\label{CDt}
\begin{split}
& C(f(V_{t_n}))_t \\
&= e^{-r(t_n -t)}\int^{\infty}_{-\infty}
f(V'_t \exp \{ \tau^{\sigma, \rho} (t)+ z
\nu^{\sigma, \rho} (t) \} )
\frac{e^{-\frac{z^2}{2}}}
{\sqrt{2\pi }}dz, \\
&\hspace{4cm} (t_{k} \leq t < t_{k+1} \leq t_n ).
\end{split}
\end{equation}
In particular,  
the expression does not contain 
$ \rho $ at $t= t_{k}$:
\begin{equation}\label{CD}
\begin{split}
&C(f(V_{t_n}))_{t_{k}} = e^{-r(t_n -t_{k})} \frac{1}
{\sqrt{2\pi}}\\
& \int^{\infty}_{-\infty}f(V_{t_{k-1}} \exp (-\frac{\sigma^2}{2}(t_n - t_{k})
+r(t_n -t_{k}) + \sigma \sqrt{t_n -t_{k}}z )) 
e^{-\frac{z^2}{2 }}dz.
\end{split}
\end{equation}
\end{thm}
A proof will be given in section \ref{pf-eqg}. 

\section{Evaluation of Transparency}
\subsection{Default Structure and the Market Value 
of the Debt}\label{DS}
As we have stated in Introduction, 
for the structure of the default we
rely 
the Merton's classical framework, 
but actually we need a deeper consideration 
on the default structure and the firm value.

Let $ \{ \delta_t: t \in \mathbf{T} \} $
be the debt structure of a firm. 
Here we assume that the maturity of each debt
is always in $ \mathbf{T} $. 
We also assume that the default occurs
only when and whenever the firm value 
$ V_{t_k} $ is less than $ \delta_{t_k} $. 
In particular, the default occurs only
at the dates of accounting report $ \mathbf{T} $.
Just like Merton's model, we can set the pay-off of the debt 
to be
\begin{eqnarray*}
\min(\delta_{t_k},V_{t_k-}), 
\end{eqnarray*}
and therefore the market value of the debt
is given by
\begin{eqnarray*}
D^{\rho, t_k}_t:= e^{-r(t_k -t)}
\mathbf{E}^\mathbf{Q} [\min(\delta_{t_k},V_{t_k-}) 
\prod_{l<k} 1_{ \{ V_{t_l-} > \delta_{t_l} \}} |
\mathcal{G}_t], \quad (t < t_k).
\end{eqnarray*}

In this debt structure, it may be natural to assume
\begin{equation*}
V_{t_k} = V_{t_k -} - \delta_{t_k},
\end{equation*}
which is somehow inconsistent with (\ref{FV1}). 
To avoid this inconsistency, we only consider 
the debt with nearest maturity: $ \delta_{t_k} > 0 $ 
and $ \delta_{t_l} = 0 $ for $ l < k $. 
under these structural assumptions, 
we may consider the market value of the firm 
as the total market value of the issued stock. 

Furthermore, since default will never occur 
before the nearest maturity $ t_k $,
we obtain the following explicit formulas
using Theorem \ref{eqg}:
\begin{equation}\label{CDEt}
\begin{split}
D^{\rho, t_k}_t 
&= V'_t \Phi(\alpha(t))\\
&- \delta_{t_k}
e^{-r(t_k-t)}\Phi(\alpha(t) + \nu^{\sigma, \rho}(t))\\
&+\delta_{t_k} e^{-r(t_k-t)}, \quad(t_{k-1} \leq t < t_k)
\end{split}
\end{equation}
where
\begin{eqnarray*}
\Phi(y):=\int^{y}_{-\infty} \frac{1}{\sqrt{2\pi}}e^{\frac{1}{2}x^2}dx,
\end{eqnarray*}
\begin{eqnarray*}
\alpha(t):=\frac{\log \frac{\delta_{t_k}}{V'_t} 
-\tau^{\sigma,\rho}(t)}
{\nu^{\sigma, \rho}(t)},
\end{eqnarray*}
and when $ t \in \mathbf{T} $ in particular, 
\begin{equation}\label{CDE}
\begin{split}
D^{\rho, t_k}_{t} 
&=V'_{t}\Phi(\beta)\\
&-\delta_{t_k}e^{-r(t_k-t)}
\Phi(\beta + \sigma \sqrt{t_k-t})\\
& +\delta_{t_k}e^{-r(t_k-t)}
\end{split} 
\end{equation}
where
\begin{eqnarray*}
\beta:=\frac{\log \frac{\delta_{t_k}}{V_{t_{k-1}}}
-(r+\frac{\sigma^2}{2})(t_k-t_{k-1})}{\sigma \sqrt{t_k-t_{k-1}}}.
\end{eqnarray*}

\subsection{Discussion for Calibration}\label{Calib}
Looking at the explicit formulas
(\ref{CDEt}) and (\ref{CDE}), we notice that
they depend on the {\em unknown} parameters
$ \sigma $ and $ \rho $ but independent of
$ \mu $. 
Furthermore, since the formula (\ref{CDE})
does not explicitly depend on $ \rho $
(of course $ V' $ does depend on $ \rho $
but this does not matter), the parameter 
$ \sigma $ can be calibrated to the market value
$ D^{\rho, t_k}_{t} $ for $ t \in \mathbf{T} $. 

Once we know the parameter $ \sigma $, 
the formula (\ref{CDEt}) contains 
only one unknown parameter $ \rho $.
Therefore, it can be calibrated to the market value
$ D^{\rho, t_k}_{t}$ for $ t \not\in \mathbf{T} $.  

A detailed guidance of the procedure 
is presented in \cite{KM}. 

\section{Proofs and  Mathematical Results}
\subsection{A Proof of Theorem \ref{firstresult}}\label{pf-first}
For $t_k \leq t < t_{k+1}$
\begin{eqnarray*}
V'_t &=& Z_t^{-1} 
\mathbf{E} [ Z_{t_{k+1} } V_{t_{k+1}} | \mathcal{G}_t ]\\
&=& Z_t^{-1} 
\mathbf{E} [ \mathbf{E}[Z_{t_{k+1} } V_{t_{k+1}} 
| \mathcal{G}_{t_k-}] | \mathcal{G}_t ]\\
&=& Z_t^{-1} 
\mathbf{E} [ Z_{t_{k+1} } V'_{t_{k+1}} | \mathcal{G}_t ].
\end{eqnarray*}
By definition we have
\begin{eqnarray*}
Z_{t_k}&=&Z_t 
\exp{\theta(W'_{t_k}-W'_t) - \frac{1}{2}}\theta^2(t_k-t)-r(t_k-t),\\
V'_{t_k}&=&V'_t
 \exp \{(\mu - \frac{(\sigma \rho)^2}{2})(t-t_k) 
 + \sigma \rho (W'_{t_k}-W'_t) \},
\end{eqnarray*}
therefore we obtain that
\begin{eqnarray*}
Z_{t_k} V'_{t_k} = Z_t V'_t 
\exp \{(\theta + \sigma \rho)(W'_{t_k}-W'_t) -
\frac{1}{2}(\theta + \sigma \rho)^2 (t_k-t) \}
\end{eqnarray*}
By the exponential martingale property, we obtain that
\begin{eqnarray*}
V'_t &=&\mathbf{E} [V_t| \mathcal{G}_t ].
\end{eqnarray*}
\qed

\subsection{A Proof of Proposition \ref{vo}}\label{pf-vo}
For $t_k \leq t < t_{k+1}$
\begin{eqnarray*}
\mathbf{E}^{Q}[K^{*}|\mathcal{G}_t]
&=&\mathbf{E}[\frac{Z_{t_k} e^{r{t_k}}}
{Z_t e^{rt}} K^{*}|\mathcal{G}_t]
=\mathbf{E}[\frac{Z_{t_k} e^{r{t_k}}}
{Z_t e^{rt}} \mathbf{E}[X|\mathcal{G}_{t_k}]|\mathcal{G}_t] \\
&=& \mathbf{E}[ \mathbf{E}[\frac{Z_{t_k}e^{r{t_k}}}
{Z_t e^{rt}} X|\mathcal{G}_{t_k}]|\mathcal{G}_t] =
\mathbf{E}[\frac{Z_{t_k}e^{r_{t_K}}}{Z_t e^{rt}}X|\mathcal{G}_t]
=\mathbf{E}^{Q}[X|\mathcal{G}_t].
\end{eqnarray*}
\qed
\subsection{A Proof of Theorem \ref{eqg}}\label{pf-eqg}

We calculate 
$C(f(V_{t_n}))_t$ for $t_{k} \leq t < t_{k+1} \leq t_n $; 
\begin{eqnarray*}
C(f(V_{t_n}))_t
=\mathbf{E^{Q}}[e^{-r({t_n}-t)}f(V_{t_n}) | \mathcal{G}_t ],
\end{eqnarray*}
where $\mathbf{E^{Q}}$ denotes the expectation 
with respect to equivalent martingale measure $ \mathbf{Q} $. 

Let us decompose $ W $ as 
\begin{equation*}
W = \rho W' + \sqrt{1 - \rho^2 } W'',
\end{equation*}
where $ W'' $ is a one dimensional Wiener process
independent of $ W' $. 
Then we have
\begin{equation*}
\begin{split}
V'_{t_k} & = V'_t \exp \big( 
\sigma \rho (W'_{t_n} - W'_t) + \sigma \sqrt{1 - \rho^2} 
(W''_{t_n} - W''_{t_k} ) \\
&\qquad -
\frac{1}{2} \sigma^2 \{ (t_n -t_k) -\rho^2 (t- t_k) \}
+ \mu (t_n -t) 
\big) \\
& = V'_t \exp \big( 
\sigma \rho \{ (W'_{t_n} - W'_t) - \theta (t_n-t) \}
+ \sigma \sqrt{1 - \rho^2} 
(W''_{t_n} - W''_{t_k} ) \\
& \qquad - 
\frac{1}{2} ( \nu^{\sigma, \rho} (t) )^2
+ r (t_n -t) 
\big) \\
&=  V'_t \exp \big( 
\sigma \rho \{ (W'_{t_n} - W'_t) - \theta (t_n-t) \}
+ \sigma \sqrt{1 - \rho^2} 
(W''_{t_n} - W''_{t_k} ) + \tau^{\sigma, \rho} (t) 
\big). 
\end{split}
\end{equation*}
Since $ \widetilde{W}' := W'_t - \theta t $ 
is a Wiener process under $ \mathbf{Q} $,
we have
\begin{eqnarray*}
& & C(f(V_{t_n}))_t = e^{-r(t_n -t)} \\
& & \mathbf{E^Q}
[f \big( V'_t \exp \{ 
\sigma \rho  (\widetilde{W}'_{t_n} - \widetilde{W}'_t)
+ \sigma \sqrt{1 - \rho^2} 
(W''_{t_n} - W''_{t_k} ) +\tau^{\sigma, \rho} (t) \} \big) 
|\mathcal{G}_t] \\
&=& \frac{e^{-r(t_n -t)}}{\sqrt{2 \pi}} 
\int_\mathbf{R} f \big( V'_t \exp \{ 
\nu^{\sigma, \rho} (t) z +\tau^{\sigma, \rho} (t) \} \big)
e^{- \frac{z^2}{2^{\,}}} \,dz .
\end{eqnarray*}
The last identity holds because 
\begin{equation*}
\sigma \rho (\widetilde{W}'_{t_n} - \widetilde{W}'_t)
+ \sigma \sqrt{1 - \rho^2} 
(W''_{t_n} - W''_{t_k} )
\end{equation*}
is independent of $ \mathcal{G}_t $
and distributed as $ N ( 0, (\nu^{\sigma, \rho} (t))^2 ) $. 
\qed

\end{document}